\documentstyle[prb,aps,multicol]{revtex} \tighten \begin{document}
\draft
\title{Screening of a macroion by multivalent ions: 
Correlation induced inversion of charge.}
 \author
{B. I. Shklovskii} \address{Theoretical
Physics
Institute, University of Minnesota, 116 Church
St. Southeast, Minneapolis, Minnesota 55455} \maketitle 
\begin{abstract}
Screening of a strongly charged macroion by 
multivalent counterions is considered.
It is shown that counterions form a strongly correlated liquid
at the surface of the macroion. Cohesive energy
of this liquid leads to additional attraction of counterions to 
the surface which is absent in 
conventional solutions of Poisson-Boltzmann equation.
Away from the surface this attraction can be taken 
into account by a new boundary
condition for the concentration of counterions near the surface.
Poisson-Boltzmann equation is solved with this 
boundary condition for a
charged flat surface, a cylinder and a sphere. In all three cases,
screening is much stronger than in the conventional approach.
At some critical exponentially small concentration of multivalent
counterions in the solution 
they totally neutralize the surface charge at small
distances from the surface. At larger 
concentrations they invert the sign
the net macroion charge. 
Absolute value of the inverted charge density can be as 
large as 20\% of that of the bare one. 
In particular, for a cylindrical macroion, 
it is shown that for screening by multivalent counterions
predictions of the Onsager-Manning theory are quantitatively incorrect.
The net charge density of the cylinder is smaller than 
their theory predicts and inverts sign with growing concentration of 
counterions. Moreover the condensation looses its universality
and the net charge linear density depends on the bare one. 
\end{abstract}

\pacs{PACS numbers: 77.84.Jd, 61.20.Qg, 61.25Hq} \begin{multicols}{2}

\section {Introduction}

Many objects with a much larger size than atomic 
are strongly charged in a water solution and are called macroions.
One can think about a rigid polyelectrolyte which, in 
a water solution, dissociates
into cylindrical macroion and monovalent small ions. 
DNA and actin are the best known examples of such
biological polyelectrolytes.
Other important types of macroions are charged lipid membranes
and charged spherical colloidal particles.
Macroions are screened by smaller ions of the solution of both signs. 
A correct description of the screening of macroions
is tremendously important for a calculation of
properties of individual macroions, for
example, the effective charge or the bending rigidity.
Screening also determines
forces acting between macroions and both
thermodynamic and transport properties of their solutions.

This paper examines the screening 
of a rigid macroion with a fixed 
and uniform distribution of charge on its surface.
Three standard geometries 
are considered below $-$ an infinite flat surface,
an infinite cylinder
and a small sphere $-$ each uniformly charged with the surface density
 $- \sigma <0$. The standard approach
for a description of such problems is the
Poisson-Boltzmann equation (PBE) 
for the selfconsistent electrostatic potential $\psi({\bf r})$
\begin{equation}
\nabla^{2} \psi = -{{4 \pi e}\over{D}} \sum Z_{i}N_{0i}
\exp\left(-{{Z_{i}e\psi}\over{k_BT}}\right).
\label{PB}
\end{equation}
Here $e$ is the charge of a proton, $D \simeq 80$ is the dielectric constant
of water, $Z_{i}e$ is the charge
of a small ion of sort $i$ and $N_{0i}$ is their concentration at the
point where $\psi=0$. The number of papers using the analytical and
numerical solutions of Eq.~(\ref{PB}) is extremely large~\cite{Frank}.
On the other hand there is an understanding that Eq.~(\ref{PB}) neglects
ion-ion correlations and is not exact.
Deviations from the distribution of charge predicted by PBE were
demonstrated numerically~\cite{Gulbrand,Roland} for the following
problem. Consider screening of a charged
surface, $x=0$, of a membrane or a film by a water solution occupying
halfspace $x>0$. Assume that there is only one sort
of counterions with the
charge $Ze > 0$ and their concentration $N(x)
= N_{0}\exp(-Ze\psi/k_BT) \rightarrow 0$ at $x \rightarrow \infty$. In
this case the solution of Eq.~(\ref{PB})
is very simple and has the Gouy-Chapman form
\begin{equation}
\label{GC}
N(x) = {1\over{2 \pi l}}{1\over {(\lambda + x)^2}},\\
\end{equation}
where $\lambda = Ze/(2 \pi l \sigma)$ is the 
Gouy-Chapman length, $l=Z^2 l_{B}$
and $l_{B} = e^{2}/(Dk_B T) \simeq 0.7$ nm is the Bjerrum length. 
At large $Z$ and $\sigma$, the length $\lambda$
can become of the order of the size 
of the water molecule or even smaller.
For example, at $Z=3$ and $\sigma = 1.0~e/$nm$^{2}$
$\lambda = 0.08$ nm.
This means that almost all ions are located in the 
first molecular layer
at the surface or, in other words, they condense
at the very surface of macroion.
This raises questions about the role of their lateral correlations
and the validity of the solution Eq.~(\ref{GC}).

It was found by numerical methods~\cite{Roland}
that for a typical charge density $\sigma$ 
deviations from Eq.~(\ref{GC})
are not large for monovalent
counterions, but they strongly increase with the charge of counterions
$Z$. It was suggested in
Ref.~\onlinecite{Rouzina96,Bruinsma,Levin,Shklov98}
that at $Z \geq 2$
repulsion between multivalent counterions condensed at the surface
is so strong that they form a two-dimensional
strongly correlated liquid (SCL) in which the short order of counterions
is similar to that of a Wigner crystal (WC).
This idea was used to demonstrate that two charged surfaces in the
presence of multivalent counterions can attract
each other at small distances.

A theory of the influence of SCL of multivalent 
counterions on the density of screening
atmosphere of a macroion has been suggested recently by
Perel and Shklovskii~\cite{Shklov99}(PS).
The PS's main idea is to treat separately two subsystems: 
two-dimensional SCL of multivalent counterions at the very 
surface and their
gas-like dilute phase at some distance 
to the right of the surface.
In the SCL, PS explicitly take into account strong 
correlations using the energy of WC as a simple 
approximation for the free energy of SCL.
On the other hand the gas-like phase is treated
in the PBE approximation, while the 
effect of SCL is taken into account with the help of a new
bounadary condition for PBE. 

In this article, PS theory is developed  in several directions. 
First, the phenomenon of charge inversion is studied 
in greater detail and the 
inverted charge as function of the counterion concentration is found.
In particular, the maximum possible 
value of the inverted charge is estimated. Second, 
PS approach is generalized to a spherical macroion. Third, I add
the comprehensive discussion of approximations made in this theory.

The next section starts with a review of the  
thermodynamic properties of a two-dimensional 
SCL, which were obtained by Monte-Carlo and other numerical methods.
It is shown that for $Z \geq 2$ and
typical charge density $\sigma$, SCL is characterized
by a large negative chemical potential of ions. In other words,
due to their lateral two-dimensional correlations ions are more 
strongly bound to the surface than
in PBE approximation. This phenomenon can be
understood as the attraction of a $Z$-valent counterion
to its correlation hole in SCL. While  PBE fails to describe these
correlation effects at the surface, it works well
at a distance from the surface, where the energy of attraction to the
correlation hole
is smaller than $k_BT$ and also $N(x)$ is small enough that
three-dimensional
correlations are very weak. It is found below, 
that PBE becomes valid at
$x \gg l/4 \gg \lambda$ and that characteristic scales of
PBE solution $\Lambda \gg l/4$.
Thus  SCL together with the
intermediate boundary layer, $\lambda < x < l/4$,
from the point of view of PBE, provide only
a new boundary condition $N(x)=N(0)$ at $x=0$ 
for the concentration $N(x)$ of multivalent counterions.
It is derived in Sec. III from the condition of
equilibrium of the gas-like phase with SCL.
Due to the large negative chemical potential of SCL, the new boundary
condition requires that $N(0)$ is
exponentially small in the dilute phase.
The section III also duscusses the structure of the intermediate
layer $\lambda < x < l/4$ between SCL
and the dilute phase, where exponential decay of $N(x)$ actually
takes place.

In Secs. IV-VI, PBE is solved with the boundary condition for $N(0)$
for the standard problems of screening of a
charged flat surface, a cylinder and a sphere
for different salt compositions of the bulk solution.
In planar geometry and for the bulk concentration of $Z:1$ salt 
$N(\infty)= 0$ I found that at $ x > l/4$
$N(x)$ obeys Eq.~(\ref{GC}) in which $\lambda$ is
replaced by exponentially large length $\Lambda$. At finite
$N(\infty)$ consequences of correlation induced
binding of counterions to the surface become even stronger.
Even at exponentially small $N(\infty) = N(0)$ the positive charge
density of
SCL completely compensates $-\sigma$. At $N(\infty) > N(0)$
it becomes even larger than $\sigma$, so that total 
charge of the surface
becomes positive. This phenomenon is called charge inversion.
First, it was noticed in numerical calculations~\cite{Roland}.
This paper presents a theory of charge inversion for the
case of screening by small size ions,  
which is totally based on correlation effects.
Recently a number of publications discussed similar phenomenon
for screening of macroions by charged
polymers~\cite{Mateescu,Gelbart,Joanny,Gurovitch,Golestanian}.

In the case of a cylinder, the conventional picture
of nonlinear screening called the Onsager-Manning
condensation should be strongly modified when
dealing with multivalent ions. Consider a cylinder with
a negative linear charge density $ - \eta$ and assume that 
$\eta > \eta_c$, where $\eta_c = Ze/l$.
Onsager-Manning theory~\cite{Manning},
confirmed by the solution of PB equation~\cite{Zimm},
shows that such a strongly charged cylinder
is partially screened by counterions residing at 
its surface, so that net (total) linear charge density
of the cylinder, $\eta^*$ is equal to the negative universal value 
$-\eta_c$. The rest of the charge is screened at much larger distances
according to linear Debye-H\"{u}ckel theory.

The Onsager-Manning condensation does 
not take into account lateral correlations of counterions.
In Sec.V  the role of these correlations is considered and 
an analytical expression for
$\eta^*$ as a function 
of $ - \eta$ and concentration $N(\infty)$ of $Z:1$ salt
is derived (see Eq.~(\ref{etaapp})). It is shown that 
due to additional binding of multivalent counterions provided 
by their SCL on the surface of cylinder, the absolute value of
the negative net charge density, 
$\eta^*$, is smaller than in the Onsager-Manning theory. 
Moreover it strongly depends on a bare linear density, $ - \eta$,
so that attractive universality of the Onsager-Manning theory
is destroyed. When concentration of counterions in the bulk
$N(\infty)$ grows, the net charge density, $\eta^*$ changes sign
from negative to positive at the point where $N(\infty) = N(0)$.
Thus, charge inversion takes place 
for a cylindrical geometry, too. 
Positive $\eta^*$ continues to grow 
with $N(\infty)$ until it reaches critical
value, $e/l_B$, for the Onsager-Manning condensation 
of monovalent negative ions.

Finally, this paper studies screening
of an uniformly charged small sphere with a negative
charge $-Q$ and radius, $a$. 
For a strongly charged sphere the solution 
of PBE is well known
~\cite{Gueron,Chaikin,Cates} and is approximately 
valid for monovalent counterions.
It shows that, in contrast with a charged plane or cylinder,
a sphere has no condensed counterions, if $N(\infty)= 0$.
This happens because the potential energy of a 
counterion on the surface 
of totally ionized sphere, $ - QZe/Da$, is finite.
At $N(\infty) > N_c$, 
where $N_c \propto \exp(- Q Z l_B/ae)$ is an 
extremely small concentration,
a fraction of positive screening charge condenses 
at the surface of the sphere and 
partially compensates its charge, so that the
net charge of the sphere, $Q^*$, 
changes in the range $0 > Q^* > -Q$.
In this regime, $Q^*$ does not depend on $Q$. This 
universality is similar to the one of the
Onsager-Manning theory. 
The rest of the screening atmosphere can be described in 
Debye-H\"{u}ckel approximation. When $N(\infty)$ 
grows, $Q^*$ becomes smaller in absolute value 
but remains negative. 

It is shown in Sec. VI that in the case of screening 
by multivalent counterions due to additional binding 
by SCL, the net charge, 
$Q^*$, behaves differently.
When $N(\infty) < N_c$, all counterions are still lost.
But at $N(0) > N(\infty) > N_c$ a larger amount of counterions 
condense at the sphere than PBE predicts.
As a result at $N(\infty) = N(0)$ the net charge, $Q^*$, 
vanishes. At $N(\infty) \gg N(0)$ 
positive $Q^*$ continues to grow and saturates at the value
$Q^* \simeq \sqrt{QZe}$. 
It should be emphasized that when correlations 
are taken into account the above mentioned universality
disappears and $Q^*$ becomes a function of the bare charge, $Q$.

Note that the net linear charge density $\eta^*$ of a cylinder
and the net charge, $Q^*$, of a sphere are measurable quantities.
In both cases they include only counterions
which binding energy exceeds $k_BT$.  These counterions move 
together with a cylinder or sphere, for example, in the electric
field. Therefore, $\eta^*$ and $Q^*$ can be studied 
in an electrophoresis experiment. In the case of charge inversion,   
a cylinder, sphere and any other macroion 
should drift in an anomalous direction.

In Sec. VII approximations of this theory  are discussed.
In the Conclusion several possible 
extensiones of this theory are mentioned.

\section{Wigner crystal and strongly correlated liquid}

It is shown below that for
$\sigma \geq 1~e$ nm$^{-2}$ and $Z \geq 2$, 
almost all charge of the plane
is compensated by SCL of counterions at its surface, which has a
two-dimensional concentration $ n = \sigma / Ze$.
In this section I discuss thermodynamic
properties of this two-dimensional system.
The minimum of the Coulomb energy of
counterion mutial repulsion and their attraction
to the background is provided by triangular close packed WC of
counterions.
Let us write energy per unit
surface area of WC at $T=0$ as $E=n \varepsilon(n)$
where $\varepsilon(n)$ is the energy per ion. One can estimate
$\varepsilon(n)$ as the interaction
energy of an ion with its Wigner-Seitz cell
of the background charge (a hexagon of
background with charge $-Ze$ and counterion in the center, 
which can be
approximately viewed as a disc with radius
$R = (\pi n)^{-1/2}$). This estimate gives
$\varepsilon(n)\sim -Z^{2}e^{2}/DR$. A more
accurate expression
for $\varepsilon(n)$ is~\cite{mara}
\begin{equation}
\varepsilon(n)= - \alpha n^{1/2}Z^{2}e^{2}D^{-1},
 \label{energyion}
\end{equation}
where $\alpha=1.96$.  Eq.~(\ref{energyion}) can be rewritten in
units of the room temperature thermal energy, $k_BT$,  as
\begin{equation}
\varepsilon(n) \simeq - 1.4~Z^{3/2}(\sigma $nm$^2/e)^{1/2}k_BT .
\label{estimate}
\end{equation}
The inverse dimensionless temperature of SCL is
 usually written in units
\begin{equation}
\Gamma = \frac{Z^{2}e^{2}}{R D k_BT}
 = 0.9\frac{|\varepsilon(n)|}{k_BT}
\label{Gamma}
\end{equation}
For example, at $\sigma = 1.0~e/$nm$^{2}$ and room temperature,
Eq.~(\ref{Gamma}) gives $\Gamma =1.2,~3.5,~6.4,~9.9$
at $Z=1,~2,~3,~4$. Thus, for multivalent counterions one
deals with a low temperature situation. $\Gamma$
is the large parameter of this paper theory. 
In its terms $R /\lambda \simeq 2\Gamma \gg 1$
 and $l/R \simeq \Gamma \gg 1$.
For example, at $Z=3$ and $\sigma = 1.0~e/$nm$^{2}$ lengths $\lambda,
R$ and $l$ are equal to 0.08,~1.0,~6.3 nm
respectively.
The small value of $\lambda$ means that almost all counterions are
located in the first molecular layer at the surface and
literally form a two-dimensional system.

It is known, however, that, due to the small shear
modulus, WC melts at a very low temperature~\cite{Gann}
near $\Gamma \simeq 130$. Nevertheless, the disappearance of
the long range order only slightly changes thermodynamic 
properties of the system.
They are determined by the short range order, which,
in the range $5 < \Gamma < 15$,
should not be significantly different
from that of the WC~\cite{Rouzina96,Bruinsma,Shklov98,Shklov99}.
This statement is confirmed by numerical
calculations~\cite{Totsuji,Lado,Gann} of thermodynamic
properties of the two-dimensional SCL 
of Coulomb particles on the neutralizing
background or so-called one-component plasma.
In the large range,
 $0.05 < \Gamma < 5000$, the excess internal
 energy of SCL per counterion (difference between internal energy 
and energy of the ideal gas with the same 
concentration),
$\varepsilon(n,T) = k_BT f(\Gamma)$,
was fitted by the expression~\cite{Totsuji}
\begin{equation}
f(\Gamma) = - 1.1 \Gamma + 0.58 \Gamma^{1/4} - 0.26
\label{intenergy}
\end{equation}
with an error less than 8\% (less than 2\% in the 
range $0.5 < \Gamma < 5000$).
The first term on the right side of
Eq.~(\ref{intenergy}) dominates at large $\Gamma$ and
leads to Eq.~(\ref{energyion}).
The other two terms provide a relatively small correction to the 
energy of WC. It is equal to 11\% at
 $\Gamma = 5$ and to 5\% at $\Gamma = 15$. 
The reason for a such small correction is that short 
range order in SCL is similar to that of WC.
For the free energy of unit area, $F$, one can write
\begin{equation}
F=F(\Gamma=0.05) + nk_BT\int_{0.05}^{\Gamma}
f(\Gamma^{\prime}){d\Gamma^{\prime}}/\Gamma^{\prime},
\label{freeenergy}
\end{equation}
so that for the chemical potential which is used below to describe
the equilibrium of SCL with the gas-like phase one obtains
\begin{eqnarray}
\mu(n,T) &=&  - k_BT\ln (n_w/n) + \mu_{s} + \mu_c(n,T),
\label{sumchem} \\
\mu_c(n,T) &=&  - k_BT (1.65 \Gamma
  - 2.61 \Gamma^{1/4} + 0.26 \ln\Gamma +1.95).
\label{cpot}
\end{eqnarray}
Here $\mu_c$ is contribution of correlations
to the chemical potential. The high temperature $\mu(\Gamma=0.05)$
with sufficient accuracy is replaced by the 
chemical potential $- k_BT\ln (n_w/n) + \mu_{s}$ 
of an ideal two-dimensional solution of ions 
in the surface layer of water with 
a two-dimensional concentration $n_w$.
The term $\mu_{s}$ is the hydration free energy per
 ion at the surface and at $n \ll n_w$
does not depend on the concentration of ions $n$
~\cite{Landau}.

The first term of Eq.~(\ref{cpot}) corresponds to the WC picture. 
Indeed, one can find directly from Eq.~(\ref{energyion}) and 
Eq.~(\ref{Gamma}) that
\begin{equation}
\mu_{WC} = \frac{d[n \varepsilon(n)]}{dn}
 = {\frac{3}{2}}\varepsilon(n) = - 1.65~\Gamma k_BT.
\label{MUWC}
\end{equation}
At large $\Gamma$, the chemical potential $\mu_{WC}$ 
dominates in Eq.~(\ref{cpot}).
The last three terms of  $\mu_c$ give
20\% correction to the WC term at $\Gamma = 5$ 
and only 10\% correction
at $\Gamma = 15$. Thus, if necessary, 
at $5 < \Gamma < 15$ one can use $\mu_{WC}$
as a first approximation. Below $\mu_c$ 
is always calculated using the full Eq.~(\ref{cpot}).

All necessary information about two-dimensional SCL 
has been presented. It is time now to study its equilibrium
with the rest of the screening atmosphere.

\section{A new boundary condition for Poisson-Boltzmann equation}

When an ion moves away from SCL it leaves behind
its negatively charged correlation hole. 
If $U(x)$ is the correlation energy of attraction to the hole,
condition of equilibrium between SCL at $x=0$ 
and the gas-like phase at a distance $x$ can be written as
\begin{equation}
\mu(n) + Ze\psi(0) = \mu(N) + Ze\psi(x) + U(x),
\label{beyond}
\end{equation}
Here $\mu(n)$ is given by  Eq.~(\ref{sumchem}),
$Ze\psi(x)$ is the counterion energy in the selfconsistent potential and
\begin{equation}
\mu(N) = - k_BT\ln N_w/N + \mu_{b}
\label{chem2}
\end{equation}
is the chemical potential of the bulk gas-like phase,
$N_w$ is the bulk concentration of
water and $\mu_{b}$ is the is the bulk hydration free energy~\cite{Landau}, which 
does not depend on $N$.
According to the terminology of Ref.\onlinecite{Landau}
Eq.~(\ref{beyond}) means that the electrochemical potential of counterions is constant.

It will be shown below that $U(x)$ becomes less than $k_BT$ at $x > l/4$.
On the other hand, in many important cases
the surface is screened so strongly that the selfconsistent 
potential changes by $k_BT$ only at 
exponentially large length, $\Lambda$, which is defined below.
Therefore, the condition of equilibrium between SCL and the
layer  $l/4 \ll x \ll \Lambda$ is
\begin{equation}
\mu(n) = \mu(N),
\label{equilibrium}
\end{equation}
Using Eq.~(\ref{sumchem}) and Eq.~(\ref{chem2}) and solving Eq.~(\ref{equilibrium})
for $N(x)$ one obtains that at $l/4 \ll x \ll \Lambda$
concentration $N(x)$ does not depend on $x$ and equals
\begin{equation}
N(0) = \frac{n}{w} \exp\left(-{{|\mu_{c}(n,T)|}\over{k_B T}}\right),
\label{bc0}
\end{equation}
where $w = (n_{w}/N_{w})\exp[(\mu_{b} - \mu_{s})/ k_BT)]$.
Below it is assumed for simplicity that $\mu_{b} = \mu_{s}$, i. e.
surface and bulk hydration free energies are equal. In this case
$w$ is the length of the order of size of the water
molecule (for estimates $w = 0.3$~nm is used below). 

The notation $N(0)$ reflects the fact that
this value plays the role of a new boundary condition 
at $x \ll \Lambda$ for important solutions of PBE which
have large characteristic length $\Lambda \gg l/4$. In such class
of solutions $N(0)$ provides an universal description of the role of SCL.
This paper deals only with this class of problems. 
Situations when one has to go beyond the 
universal boundary condition (\ref{bc0}) and start directly from
Eq.~(\ref{beyond}) will be studied in the next paper.
 
Due to the large value of
$|\mu_{c}(n,T)|$, the concentration $N(0)$ can be very small. 
For example, at $\sigma = 1.0~e/$nm$^{-2}$, and 
$Z = 2, 3$ and 4 at which $\Gamma = $3.5, 6.4 and 9.9, 
and one gets, according to Eq.~(\ref{cpot}),
$|\mu_{c}(n,T)|/k_B T =$4.5, 8.8 and 14.3 respectively.
This gives for $N(0)$ = 30~mM, 0.3~mM and 0.8~$\mu$M for $Z = 2, 3$ and 4
(1~M = 6~10$^{26}$m$^{-3}$). 
It is clear now that $|\mu_c(n,T)|$ plays 
the role similar to the work function
for thermal emission, to the free energy of chemosorption or 
to the evaporation energy for the cases of equilibrium gas-liquid 
or gas-solid interfaces. The concentration $N(0)$ 
is similar to the  density of the saturated vapor.

Thus, correlation effects in SCL provide additional strong
binding of counterions to the macroion surface.
Qualitative arguments for such binding can be found in 
Ref.~\onlinecite{Roland}.
We would like to stress that such binding does not happen at $Z=1$.
Indeed, at $\sigma = 1.0~e/$nm$^{-2}$ one 
obtains from Eq.~(\ref{Gamma}) 
and Eq.~(\ref{cpot}) that $\Gamma = 1.2$ and $\mu_c(n,T)/k_BT = 1.3$.
Therefore, the boundary condition Eq.~(\ref{bc0}) does not produce
nontrivial effects and standard 
solutions of PBE remain approximately valid.

Below, I justify the role of the 
distance $l/4$ and give an idea how $N(x)$
evolves from $n/\lambda$ at $x\sim \lambda$ to $N(0)$ at $x=l/4$.
Let us move one ion of the
crystal along the $x$ axis. As it is mentioned above, 
it leaves behind its
correlation hole. In the range
of distances $\lambda \ll x \ll R$, the correlation hole is
approximately a disc of the surface charge
with radius $R$ (the Wigner-Seitz cell) and 
the ion is attracted to the
surface by its uniform
electric field $E = 2 \pi \sigma/ D$.
Therefore, if $\lambda$ were larger than $w$ one would get
$N(x)= (n/\lambda)\exp(-x/\lambda)$
at $x \ll R$. In the cases of our interest $\lambda < w $ and
 $N = n/w$ at $x < w$, while at $w \ll x \ll R$
\begin{equation}
N(x) = \frac{n}{w}\exp(-x/\lambda).
\label{disc}
\end{equation}
At $ x \gg R$ the correlation hole radius grows and 
becomes of the order of $x$. Indeed SCL on the uniform background 
can be considered as a good conductor in the plane $(y, z)$.
It is known that a charge at a distance, $x$, from a metallic plane
attracts an opposite charge into
a disc with the radius $\sim x$ or, in other words, 
creates its point-like
image on the other side of the plane at the distance $ 2x$
from the original charge. The same thing happens to SCL. The removed ion
repels other ions of SCL and creates a correlation hole in the form of
a negative disc with the charge $-Ze$ and the radius $x$. The
correlation hole attracts the removed ion and decreases its potential
energy by the Coulomb term 
\begin{equation}
U(x) = - Z^{2}e^{2}/ 4Dx.
\label{Ux}
\end{equation}
This effect provides the correction to
the activation energy of $N(x)$ :
\begin{equation}
N(x)=\frac{n}{w}\exp\left(-{{|\mu(n)| - Z^{2}e^{2}/4 Dx}
\over{k_B T}}\right)~~( x \gg R).
\label{Nx}
\end{equation}
The similar ``image" correction to the work function of a metal is
well-known
in the theory of thermal emission. The correction decreases with
$x$ and at $x =l/4$,
becomes equal to $k_B T$, so that $N(x)$ saturates at the value $N(0)$.
The dramatic difference between the exponential 
decay of Eqs.~(\ref{disc})
and (\ref{Nx}) and the $1/x^2$ law of Eq.~(\ref{GC})
 is obviously related to the
correlation effects neglected in PBE.
Recall that it was assumed in the beginning of this paper that the
charge of the surface is almost totally compensated by SCL.
 Exponential decay of
$N(x)$ with $x$ confirms this assumption and at 
$\Gamma \gg 1$ makes this
theory self consistent.

Consider now what happens with $N(x)$ at distances $x \gg l/4$.
At such distances, correlations of the removed ion  
with its correlation
hole in SCL are not important and the correlation between
ions of the gas phase are even weaker
because $N(x)$ is exponentially small. Therefore, 
one can return to PBE. In the next
section  solutions of PBE for the planar geometry for
different concentration of salt are discussed.

\section{Planar geometry}

The solution of PBE with the boundary
condition (\ref{bc0}) and $N(\infty)= 0$ is similar to Eq.~(\ref{GC}):
\begin{equation}
N(x) = {1\over{2 \pi l}}{1\over {(\Lambda + x)^2}}~~~~~~(x \gg l/4),
\label{GCnew}
\end{equation}
where the new renormalized Gouy-Chapman length, $\Lambda$, is
exponentially large
\begin{equation}
\Lambda = (2\pi l N(0))^{-1/2}= \sqrt {\frac {w}{2 \pi n l}}
\exp\left({{|\mu_{c}(n,T)|}\over{2k_B T}}\right).
\label{Lambda}
\end{equation}
For example, at $\sigma = 1.0~e/$nm$^{-2}$, Eq.~(\ref{Lambda})
gives $\Lambda \simeq 1.8, 12.3, 166$ nm at $Z=2,3,4$. 
These lengths should be compared with 
$l/4 = 0.7, 1.6, 2.8$~nm respectively. We see that $\Lambda \gg l/4$ 
for $Z \geq 2$. This justifies the use
of Eq.~(\ref{bc0}) as the boundary condition for the large distance
solution of PBE. 

Using Eq.~(\ref{GCnew}) one finds that the total
surface charge density located at distances $x < l/4$
\begin{equation}
\sigma^* = - \sqrt {N(0) /(2\pi l_{B})} = - \sigma (\lambda/\Lambda).
\label{Sigma}
\end{equation}
For $\sigma = 1.0~e/$nm$^{-2}$ one finds that
$\sigma* = 7~10^{-2}\sigma$ at $Z=2$,
 $\sigma* = 7~10^{-3}\sigma$ at $Z=3$ and
$\sigma* =4~10^{-4}\sigma$
at  $Z=4$. Corrections to $\mu_{c}(n,T)$ and
$N(0)$ related to such small $\sigma^*$ can be, of course, neglected.

One can compare these  results with predictions of Eq.~(\ref{GC}).
Integrating Eq.~(\ref{GC}) from $l/4$ to $\infty$ one finds
$\sigma* = 2Ze/\pi l^2$, i.e. $\sigma* = 5~10^{-2}~e/$nm$^{-2}$
at $Z=3$ and  $\sigma* = 2~10^{-2}~e/$nm$^{-2}$ at $Z=4$.
These values of  $\sigma*$ are much larger than 
Eq.~(\ref{Sigma}). Thus, binding to the surface is strongly 
enhanced by correlation effects.

Until now this paper has addressed the case of
 extremely dilute solution, when
$N(\infty)= 0$. Consider the case of a finite concentration, 
$N(\infty)$, of a $Z:1$ salt
in the bulk of solution, or in other words, of a 
concentration $N(\infty)$ of Z-valent counterions and concentration 
$N_{-}(\infty) = ZN(\infty)$ of
neutralizing ions with the charge $-e$. This adds the
Debye-H\"{u}ckel screening radius
\begin{equation}
r_s = \left(4\pi l N(\infty) (1+ 1/Z)\right)^{-1/2}
\label{screenrad}
\end{equation}
to the problem. If $N(\infty) \ll N(0)$ the screening radius $r_s \gg
\Lambda$, and the fact that $N(\infty)$ is finite
changes only the very tail of Eq.~(\ref{GCnew})
making the decay of $N(x)$ at $x \gg r_s $ exponential. At $x \ll r_s$,
 still
$N(x) \gg ZN_{-}(x)$ and all previous results are valid.
However, when $N(\infty)$ approaches $N(0)$,
the solution changes dramatically and $\sigma^*$ vanishes. Indeed,
when $N(\infty) = N(0)$ concentration
$N(x) = N(\infty)\exp(-{Ze\psi}/{k_BT}) =  N(0)$ stays constant and
potential $\psi(x) = 0$  at $x > l/4$.
This means that the surface is completely
neutralized at distances $0 <x <l/4$.

If $N(\infty) \gg N(0)$ negative charges dominate at $x \ll r_s$.
Indeed in the PBE approach,
\begin{eqnarray}
N(x) &=& N(\infty)\exp(-{Ze\psi}/{k_BT}), \label{pm1}\\
N_{-}(x)& =&N_{-}(\infty)\exp({e\psi}/{k_BT}) \label{pm2},
\end{eqnarray}
and when concentration $N(x)$ decreases with decreasing $x$,
the electrostatic potential,$\psi(x)$, grows
and $N_{-}(x)$ increases. One can
derive a boundary condition for $N_{-}(x)$ at $x=0$ from
Eqs.~(\ref{pm1})
and (\ref{pm2}).
For this purpose, one should first express 
$\psi(0)$ through $N(0)$ with
help of
Eq.~(\ref{pm1}), and then find $N_{-}(0)$ from  Eq.~(\ref{pm2}). This
gives
\begin{equation}
N_{-}(0) =ZN(\infty)[N(\infty)/N(0)]^{1/Z},
\label{newbc}
\end{equation}
where $N(0)$ is given by Eq.~(\ref{bc0}).
Then the solution of PBE for $N_{-}(x)$ at $x \ll r_s$
has the form similar to Eq.~(\ref{GCnew})
\begin{equation}
N_{-}(x) = {1\over{2 \pi l_{B}}}{1\over {(\Lambda_{-} + x)^2}},
\label{GCnewm}
\end{equation}
where
\begin{equation}
\Lambda_{-} = (2\pi l_{B} N_{-}(0))^{-1/2},
\label{Lambdam}
\end{equation}
and $l_{B}=e^{2}/(Dk_B T)$ is the Bjerrum length.
To compensate the bulk negative
charge the positive surface charge density of SCL becomes
larger than $\sigma$, so that the net surface
charge density $\sigma^*$ becomes positive.
Similarly to Eq.~(\ref{Sigma}), it is
\begin{equation}
\sigma^* = e~\sqrt \frac{N_{-}(0)}{2\pi l_{B}}
= \frac{e}{2\pi l_{B}\Lambda_{-}}.
\label{Sigmaover}
\end{equation}
This phenomenon is called charge inversion and is, of course,
impossible in the framework of the
standard PBE. Technically, charge
inversion follows from the small value of $N(0)$ in Eq.~(\ref{bc0}).
Its physics is related to the strong binding of counterions
at the charged surface
due to the formation of SCL. Remarkably,
when $\Gamma \gg 1$, this
phenomenon happens under the influence of a
very small concentration of salt.

According to Eqs.~(\ref{Sigmaover}) and (\ref{newbc}) the net density $\sigma^*$
continues to
grow with $N(\infty)$ at $N(\infty) \gg N(0)$.
It is interesting to study how far it can grow and 
how strong the charge inversion can be. The use of PBE with the boundary conditions
Eqs.~(\ref{bc0}) and (\ref{newbc}) is valid if $\Lambda_{-} > l/4$.

To estimate the maximum value of $\sigma^*$,
which can be reached within the range of validity of this theory
one can substitute 
$\Lambda_{-} = l/4$ into Eq.~(\ref{Sigmaover}). This gives
\begin{equation}
\sigma^* \simeq \frac{2e}{\pi l_{B} l} = \sigma \frac{2R^2}{Zl_{B}^2}.
\label{Sigmaovermax}
\end{equation}
For example, at $Z=3$ and $\sigma = 1.0~e/$nm$^{-2}$ one obtains
$\sigma^* \simeq 0.15 \sigma$. 
To find  $\sigma^*$ as a function of $N(\infty)$ in the whole 
range where $\Lambda_{-} \gg l/4$, one should solve
Eq.~(\ref{Sigmaover}) selfconsistently
substituting $n= (\sigma + \sigma^*)/Ze$ into
Eqs.~(\ref{bc0}), (\ref{newbc}) and (\ref{Lambdam}).

One can show that at $R < \Lambda_{-} < l/4$, when our theory based on the
universal boundary condition Eq.~(\ref{bc0}) is not valid, $\sigma^*$ continues to grow. 
If $\Lambda_{-}$ becomes smaller than the radius of a Wigner-Seitz cell $R$,
negative ions screen each counterion  separately. The effective
charge of counterions becomes smaller than $Z$. This weakens their lateral
interactions and makes $N(0)$ larger. Therefore, $\sigma^*$ starts to decrease. 
The maximum value of $\sigma^*$ is reached at $\Lambda_{-} \simeq R$ and
is close to $e / (2\pi l_{B}R)$. For $Z=3$ and $\sigma = 1.0~e/$nm$^{-2}$ 
this gives $\sigma^* \simeq 0.24 \sigma$.

I will not try here to make the above estimates of
the maximum $\sigma^*$ more accurate
because of sensitivity of this estimate to the ion
size due to the following reason. It was assumed above
that when $\sigma^*_{max}$ is reached, all salt molecules are still fully
dissociated in water, so that the concentration, $N(\infty)$
of ions with charge $Z$ is equal to the concentration
of the salt, $N_{s}(\infty)$.
In reality, at very large $N_{s}(\infty)$,
the concentration of fully ionized counterions, $N(x)$,
saturates at the level
\begin{equation}
N_{max}(\infty) \sim b^{-3}\exp( - Ze^{2}/bDk_B T),
\label{Nmax}
\end{equation}
where $ -Ze^{2}/bD$ is the Coulomb
interaction energy of positive $Z$-valent ion
with negative monovalent one at the minimum distance between them, $b$.
In this case, the majority of counterions keep a negative ion.
One can refer to such a complex as $(Z-1)$-ion.
The transition to such regime happens when the concentration of salt,
$N_{s}(\infty)$, reaches $N_{max}(\infty)$.

Substituting $N_{max}(\infty)$ into Eq.~(\ref{GCnewm})
and then Eq.~(\ref{GCnewm}) into Eq.~(\ref{Sigmaover})
one finds that at $Z=3$, $\sigma = 1.0~e/$nm$^{-2}$ and $b \geq 0.4$ nm
this limitation
of dissociation is not important. For smaller $b$
charge density, $\sigma^*$,
saturates at the value $N_{max}(\infty)$ which is smaller
than Eq.~(\ref{Nmax}), and stays at
this level until the concentration of $(Z-1)$-ions becomes so large
that they replace fully ionized ions at the surface.
This leads to the drop of $\sigma^*$.

Note once more that dramatic changes of the
screening atmosphere described above do not happen at $Z=1$
when $\Gamma \sim 1$ and $|\mu_c(n,T)|/k_BT \sim 1$.
The standard Gouy-Chapman solution of the PBE,
Eq.~(\ref{GC}), remains valid in this case.

\section{Screening of uniformly charged cylinder}

Consider screening of an infinite rigid cylinder
with a radius $a$, a negative surface charge density
$-\sigma$ or, in other words, with a negative linear 
charge density $ -\eta = -2\pi a \sigma$. Assume that $\sigma$ is large enough so that
the surface of the cylinder
is covered by a two-dimensional SCL with $R < 2\pi a $ and with
$\Gamma \gg 1$.
Such a cylinder can be a first order approximation for the
double helix DNA, where $a =1~$nm, $\eta = 5.9~e/$nm, $\sigma
=0.94~e/$nm$^{2}$, and for $Z = 3$ radius of the 
Wigner-Seitz cell $R\simeq 1$~nm and $l=6.3~$nm.

A screening atmosphere of a cylinder is described by the concentration
$N(r)$, where $r$ is the distance from the cylinder axis.
The solution of PBE is known~\cite{Zimm,Frank}
to confirm the main features of the famous
Onsager-Manning~\cite{Manning} picture of the counterion condensation.
This solution depends on the
relation between $\eta$ and $\eta_c= Ze/l =k_BT/eZD$. For a weakly
charged cylinder with $\eta \ll \eta_c$,
the screening is linear and 
can be described by the  Debye-H\"{u}ckel
approximation. For $\eta > \eta_c$, screening
becomes nonlinear and most of the screening charge, 
$-(\eta - \eta_c)$, is
located at the cylinder surface, while  at $N(\infty)=0$ the rest of 
the screening charge,
$ \eta_c$ is spread in the bulk of the solution.
This means that at large distances, the net charge density
of the cylinder, $\eta^*$, equals $- \eta_c$ and does not depend on
$\eta$. Note that this is
different from the planar geometry where all the charge
is bound to the surface, so that far enough from the surface, the
net surface density vanishes (the finite $\sigma^*$ 
was defined at $x <l/4$). At a finite $N(\infty)$, the charge density $\eta^*$
is screened only at linear screening radius $r_s$.

Here I deal with strongly charged cylinder, $\eta \gg \eta_c$.
It easy to check that this inequality follows from our assumptions
$R < 2\pi a$ and $\Gamma \gg 1$. It is also fulfilled for the case of
DNA for which $\eta/ \eta_c \simeq 4Z$. The goal in this case 
is to verify whether in the case of multivalent counterions 
elegant statements of the
Onsager-Manning theory~\cite{Manning} should be changed
due to SCL at the surface of a strongly charged cylinder.

As in the previous section, the boundary condition
Eq.~(\ref{bc0}) is used below to allow for 
additional binding of counterions
by SCL. One can introduce a radius $r_T$,
at which energy of interaction between a counterion and its
correlation hole, $U(r)$, becomes of the order of $k_BT$, so that
the boundary condition $N(r) = N(0)$ can be used. For a cylindrical
geometry, $r_T$, strictly speaking,
differs from its analog for a planar problem $l/4$. Indeed, at
$r \gg R$ energy $U(r)$ can be
calculated as energy of attraction the charge $Ze$ to an 
infinite metallic wire with the radius $a$:
\begin{eqnarray}
U(r) &=& - Z^{2}e^{2}/ 4D(r-a)~~~~~~~~(R < r-a < a),\nonumber\\
U(r) &=& - \pi Z^{2}e^{2}/4Dr\ln(r/a)~~~~~~~~( r \gg a).
\label{Ur}
\end{eqnarray}
One can find $r_T$ from the equation $|U(r_T)| = k_BT$:
\begin{eqnarray}
r_T &=& a + l/4~~~~~~~~(l/4 < a),\nonumber\\
r_T &=& \frac{ \pi l}{4\ln(l/4a)}~~~~~~~~( l/4 \gg a).
\label{rT}
\end{eqnarray}
At distances $r_T < r < r_s$ the electrostatic potential of
the linear charge density $\eta^*$ is not
screened and the boundary condition of Eq.~(\ref{bc0}) 
can be used to write
\begin{eqnarray}
N(r)&=&N(0)\exp\left(-{{Ze[\psi(r) -
\psi(r_T)]}\over{k_BT}}\right)\nonumber\\
&\simeq&N(0)\exp\left({{2\eta^*}\over{\eta_c}}\ln(r/r_T)\right).
\label{unscreenN}
\end{eqnarray}
At $ r = r_s$ concentration $N(r_s) \simeq N(\infty)$.
The solution of this equation for $\eta^*$ is
\begin{equation}
\eta^* = - \eta_c {{\ln
[N(0)/N(\infty)]}\over{\ln(r_s/r_T)^{2}}}.
\label{etaapprs}
\end{equation}
According to Eq.~(\ref{rT}), at a not very large $l/4a$ 
one can use estimate, $r_T \sim l/4$.
Substituting Eq.~(\ref{screenrad}) into Eq.~(\ref{etaapprs}), 
one arrives at
\begin{equation}
\eta^* = - \eta_c {{\ln [N(0)/N(\infty)]} \over\ln{[4/(\pi N(\infty)l^{3})]}}.
\label{etaapp}
\end{equation}
It is clear from Eq.~(\ref{etaapp}) that if two logarithms are close to
each other, i. e. if
\begin{equation}
\ln \frac{N(0)^{2}l^{3}}{N(\infty)} \gg 1
\label{OMcond}
\end{equation}
the Onsager-Manning theory is approximately correct and $\eta^*$
 approaches $- \eta_c$.
If $\Gamma \sim 1 $ and $\mu_c(n,T) \sim k_BT$ 
concentration $N(0) \sim
n/w$ is large
and inequality~(\ref{OMcond}) is fulfilled at any 
reasonable $N(\infty)$.
Thus for typical charge density $\sigma$ and $Z=1$ 
the Onsager-Manning result
is  rederived.
However,  for screening by multivalent ions  $\Gamma \gg 1 $,  in
Eq.~(\ref{bc0})
 $\mu_c(n,T) \gg k_BT$ and concentration $N(0)$ is 
exponentially small so that
values of $N(\infty)$ at which $\eta^*$ is close to
$- \eta_c$ are extremely small. For example, to get $\eta^* 
= - 0.75\eta_c$
one needs
$N(\infty) = N_{0.75} = 0.02 N(0)^{2}l^{3}$. At $\sigma 
= 1.0~e/$nm$^{-2}$
and $Z=3$ it
is shown above that $N(0)= 1.7~10^{23}$m$^{-3} = 0.3$~mM and
therefore $N_{0.75} = 2~10^{20}$m$^{-3}= 0.3~\mu$M. Switching to $Z=4$,
one has $N(0)= 5~10^{20}$m$^{-3} = 0.8~\mu$M which
results in an unrealistically small
$N_{0.75} = 2.5~10^{15}$m$^{-3}$.

On the other hand, in disagreement with the Onsager-Manning theory one
obtains from Eq.~(\ref{etaapp})
that $|\eta^*| \ll \eta_c$ when a concentration $N(\infty)$
 of the salt is still
exponentially small, namely $N(0)^{2}l^{3} \ll N(\infty) \ll N(0)$.
Moreover, according to Eq.~(\ref{etaapp}) 
$\eta^*$ vanishes at $N(\infty) =
N(0)$. This result is easy to understand 
without calculations. Indeed, in this case $N(r) =
 N(\infty)\exp(-Ze\psi(r)/k_BT) = N(0)$ stays constant
 and $\psi(r)=0$ at
all $r > l/4$,
 so that all of the charge of the polyelectrolyte is 
compensated inside cylinder with $r=l/4$.

The difference from the Onsager-Manning theory becomes even more apparent at
$N(\infty) > N(0)$
when the density $\eta^*$ becomes positive.
Note that this charge inversion takes place still at exponentially small
$N(\infty)$. A positive $\eta^*$ continues to grow with $N(\infty)$ until
it reaches critical density 
\begin{equation}
\eta^{*}_{max} = e/l_B
\label{etamax}
\end{equation}
and the standard Onsager-Manning condensation of monovalent
negative ions starts.
According to Eq.~(\ref{etaapp}) this happens at $N(\infty) = N_{sat}$, where
\begin{equation}
N_{sat} \sim l^{-3}( N(0)l^3)^{1/(Z+1)}. \label{Nsat}
\end{equation}
At $N(\infty) > N_{sat}$ charge density $\eta^*$
remains fixed at the level $e/l_B$.
Condensed negative ions eventually screen lateral interaction of
counterions in SCL, $|\mu_c(n,T)|$
decreases and $\eta^*$ drops. Comparing Eq.~(\ref{Nsat}) with
Eq.~(\ref{Nmax}) for the maximum
concentration, $N_{max}(\infty)$,
 of fully dissociated $Z:1$ salt one sees that they are
quite close, if ion size $b$ is not too small. 
For a very small distance of the closest approach $b$, the growth of
$\eta^*$ is limited earlier 
by the condition $N(\infty) < N_{max}(\infty)$.

To summarize, the net charge $\eta^*$ 
as a function of salt concentration $N(\infty)$ 
is given by Eq.~(\ref{etaapp}). It changes 
in the range $e/l_B > \eta^* > - \eta_c$ when $N(\infty)$ grows.
Strictly speaking, to quantitatively describe $\eta$ as a function of $N(\infty)$,
one should use the  
self consistent concentration, $n = (\eta + \eta^*)/2\pi a$, in
Eq.~(\ref{etaapp}) for $N(0)$.

Finally, it should be emphasized that this result does not demonstrate
the attractive universality
of the Onsager-Manning theory. According to Eq.~(\ref{etaapp}),
$\eta^*$ depends on $\eta$ through $\mu(n)$ in Eq.~(\ref{bc0}).
Thus, for the screening by multivalent ions at $\Gamma \gg 1$, and
at any reasonable $N(\infty)$ the predictions of
Refs.~\onlinecite{Manning,Zimm} are qualitatively incorrect.

Return now to the case of a small concentration of
 a $Z:1$ salt, $N(\infty) \ll
N(0)$, and consider
what happens to $\eta^*$ when a 1:1 salt with a larger concentration,
$N_1 \gg N(\infty)$, is added to the solution. This is a realistic
experimental situation.
Such a problem can be solved with the help of Eq.~(\ref{etaapprs}),
if one substitutes $r_s = (8\pi l_B N_1)^{-1/2}$ instead of
Eq.~(\ref{screenrad}). The result is
\begin{equation}
\eta^* = - \eta_c {{\ln [N(0)/N(\infty)]}
\over\ln{[2Z^{2}/(\pi N_1 l^{3})]}}.
\label{etaapp1}
\end{equation}
Eq.~(\ref{etaapp1}) shows how at a given $N(\infty) \ll N(0)$, the 
absolute value of
the net negative charge density increases with $N_1$. At large enough $N_1$,
monovalent counterions replace counterions with charge $Ze$ at the surface
of the macroion.
This replacement happens when 
corresponding change of free energy vanishes, i.e. at
\begin{eqnarray}
\mu_{c}(n,T) - k_BT\ln (n_w/n) - Zk_BT\ln (N_w/N_1) =\nonumber\\
Z\mu_{1} - Zk_BT\ln (n_w/Zn) - k_BT\ln (N_w/N(\infty)).
 \label{replacement}
\end{eqnarray}
Here the left side is the free energy of 
a $Z$-valent counterion at the surface
and $Z$ monovalent ones in the bulk, while the right side is the free
energy of $Z$ monovalent ions at the surface and a $Z$-valent ion in
the bulk. $\mu_{1}$ is the correlation part of the chemical potential of a
monovalent ion.
One can neglect $Z\mu_{1}$ in comparison with $\mu_{c}(n,T)$ because, as it was 
mentioned above,
$\mu_{1}$ is numerically small and the latter
quantity is proportional to
$Z^{3/2}$. Solving Eq.~(\ref{replacement}) for $N_1$ one finds %
\begin{equation}
N_1 = \frac{n}{w} \left( \frac {N(\infty)}{N(0)}\right)^{1/Z}.
\label{replacement1}
\end{equation}
Substituting this $N_1$ into Eq.~(\ref{etaapp1}), one sees that at the
moment of
replacement $\eta^* \simeq -\eta_c Z = - k_BT/e$, providing a natural
crossover to the case of screening by
exclusively monovalent counterions.

\section{Screening of an uniformly charged sphere}

Consider application of
this theory to a sphere with a small radius $a= 2-5$~nm and 
with a charge $-Q$ screened by $Z:1$
salt with concentration $N(\infty)$ in the bulk.
At large enough surface
charge density
$\sigma = Zen_{0} = -Q/4 \pi a^2$ the surface
is covered by SCL of $Z$-valent counterions. The goal is to find a 
concentration $n$ of
this SCL and the net charge of the sphere,
\begin{equation}
Q^* = 4\pi a^2 nZe - Q,
\label{q}
\end{equation}
as a function of $a$, $Q$, $Z$, $T$ and $N(\infty)$. In the case of a
sphere, screening
atmosphere is characterized by concentration of $Z$-valent ions, $N(r)$, as
a function
of distance $r$ from the sphere center.
For simplicity, assume that $a > l/4$ so that the boundary condition
Eq.~(\ref{bc0}) is valid
at $ r = a + l/4$ where the curvature of the sphere
can be neglected. At distances from the surface
$r - a \ll r_s $  one can neglect screening and find the
concentration, $N(r)$, similarly to
Eq.~(\ref{unscreenN}) as
\begin{eqnarray}
N(r)&=&N(0)\exp\left(-{{Ze[\psi(r) -
\psi(a + l/4)]}\over{k_BT}}\right)\nonumber\\
&\simeq&N(0)\exp\left(\frac{Q^*Ze(a^{-1} - r^{-1})}{Dk_BT}\right).
\label{unscreenNN}
\end{eqnarray}
At the distance $r= a + r_s$ where linear Debye-H\"{u}ckel theory starts to work,
one has
\begin{equation}
N(a + r_s) \simeq N(\infty).
\label{Master}
\end{equation}
Solving this equation in the case $r_s \gg a$ one obtains
\begin{equation}
Q^* = - {\frac {ae}{l_B Z}} \ln \frac {N(0)}{N(\infty)} = - {\frac {a}{l_B Z}}
\left(\ln \frac {n/w}{N(\infty)} + \frac{|\mu_{c}(n,T)|}{k_BT} \right).
\label{qq}
\end{equation}
This equation is similar to Eq.~(\ref{etaapp}). 
It is to be solved for
$Q^*$, $n$ and $|\mu_{c}(n,T)|$
together with Eq.~(\ref{q}) and Eq.~(\ref{cpot})
(or its low temperature version Eq.~(\ref{MUWC})).

In the case of monovalent counterions, when $\Gamma \sim 1 $, 
$\mu_c(n,T)/ k_BT \sim 1$, so that correlations do not play any
role in Eq.~(\ref{qq}) and solution does not differ from the solution of PBE.
In this case, the concentration $N(0) \simeq n/w  \sim n_0/w $
is much larger then any
reasonable $N(\infty)$  so that $\ln (N(0)/N(\infty)) > 0$ 
and $Q^* < 0$.
Thus, Eq.~(\ref{qq}) describes the partial compensation of charge
$-Q$ by positive charge $Q + Q^* = 4\pi a^2Ze n$ of 
counterions condensed at the very surface
of the sphere. The rest of the screening charge, $- Q^*$  
is situated at the
distance $r_s$ from the sphere
in Debye-H\"{u}ckel atmosphere. 
It is clear now that nonlinear screening of a sphere
is similar to Onsager-Manning condensation in the case of a cylinder
~\cite{Gueron,Chaikin,Cates}.
In both cases there are two separate groups
of counterions: condensed and free. Moreover, for a sphere there is a
similar universality of the net charge $Q^*$.
Indeed, when $N(\infty) \ll n_{0}/w$ the dependence of $Q^*$ on $Q$ 
is negligible.
(One can evaluate this dependence substituting $n_0$ for
 $n$ in Eq.~(\ref{qq}).)
The only qualitative difference between the screening of 
a sphere and a cylinder is that at unrealistically small
$N(\infty) \leq N_c$ where
\begin{equation}
N_c = {\frac {1}{4\pi a^{2}w}} \exp \left( - {\frac {Ql_BZ}{ae}}
 \right)
\label{Nc}
\end{equation}
the last counterion leaves the surface\cite{Zimm} and $Q^* = - Q$.

On the other hand, in the case of screening by multivalent ions, correlations 
significantly
change the above described mean field dependence of $Q^*$ on $N(\infty)$.
These changes start, however, only at $N(\infty) > N_c$, 
because without condensed counterions
correlations can not play any role.
At $N(\infty) \gg N_c$, when $n$ grows and
 becomes comparable with $n_0$
one obtains that $\Gamma \gg 1 $, $\mu_c(n,T) \gg k_BT$
 and, according to
Eq.~(\ref{bc0}), $N(0)$ is exponentially small.
Therefore, it follows from Eq.~(\ref{qq}) that the negative net charge, $Q^*$,
grows (decreases in the absolute value) 
faster than it does in the case of monovalent. Eventually, $Q^*$ vanishes
at exponentially small $N(\infty) = N(0)$.

At $N(\infty) > N(0)$ the net charge, $Q^*$, becomes positive 
and continues to grow.
As in other geometries this charge inversion happens because of strong
binding of counterions by SCL. At large enough $N(\infty)$ one
can neglect the first term in parenthesis of Eq.~(\ref{qq}).
Then using a low temperature expression
Eq.~(\ref{MUWC}) for $\mu_c(n)$
one find that $Q^*$ saturates at the positive value
\begin{equation}
Q^*_{max} = \beta \sqrt{Q Ze},
\label{qs}
\end{equation}
where $\beta = 3\alpha/4\sqrt{\pi} \simeq 0.84$.
For example, at $Z=3$ for
a sphere with bare charge $- 50e$
($Q=50e$) one arrives at the net charge $Q^*_{max} \simeq +10e$.
Eq.~(\ref{qs}) is remarkably simple and universal:
$Q^*_{max}$ does not depend on a sphere radius $a$.

Eq.~(\ref{qs}) for $Q^*_{max}$ is valid until
one of the two following events happens. First, a
concentration $N(\infty)$ can reach maximum concentration
$N_{max}(\infty)$ of fully dissociated $Z:1$
salt (see Eq.~(\ref{Nsat})).
Secondly, a condensation of monovalent negative ions on 
the positive sphere can start.
Condensed counterions eventually screen the lateral 
interaction of counterions
in SCL and effectively change their charge from 
$Z$ to $Z-1$. As a result
$|\mu_c(n,T)|$ decreases and $q$ starts to drop.
Similarly to Eq.~(\ref{Nsat}) condensation starts at
$N(\infty) \propto \exp(-\mu_c/(Z+1)k_BT)$. 
This concentrations is close to
$N_{max}(\infty)$, if ion size, $b$, is not too small. 
For a very small $b$ condition $N(\infty) \ll N_{max}(\infty)$ 
is more restrictive. For
a small sphere with radius $a = 2-5$ nm 
both restrictions start to work while $r_s \gg a$.
Therefore, I do not consider here the case $r_s < a$. 
For larger spheres, $r_s$ can become smaller 
than the above mentioned other limits on $N(\infty)$ start to work. 
In this case, the sphere effectively works as a flat surface 
and one can use results of Sec. IV.

\section{Discussion of approximations}

In this section, approximations used in this paper are discussed. 
First, it was assumed that charges at the surface 
of a macroion are fixed and can not move.
In the case of a solid or glassy surface, for example,
colloidal particles and rigid 
polyelectrolytes, such as double helix DNA, 
this approximation seems to work well.
On the other hand, for charged lipid membranes it can be violated.
If the surface charges are mobile they can accumulate
near $Z$-valent counterions,
forming short dipoles directed perpendicular to the surface.
These dipoles interact weakly with each other so that the
energy of their lateral correlations is smaller than in SCL
on the uniform background.
On the other hand, such concentration of the 
surface charge under the counterion 
by itself creates an additional binding of counterions 
to the surface. 
As a result the negative chemical potential, $\mu(n,T)$, 
becomes larger in the absolute value and the boundary concentration 
$N(0)$ becomes smaller.
The theory of this paper expresses everything through $N(0)$. 
Therefore, the unusual effects of complete compensation of macroion 
charge and of the charge inversion become stronger.

The second approximation made above is the assumption that 
fixed charge is uniformly distributed at the surface.
Localized charges are actually discrete. 
Therefore it makes sense to discus whether -$e$ charges,
for example, randomly 
distributed on the surface, work as an uniform background. 
In the limit $Z\gg 1$, the 
repulsion between $Z$-valent counterions is much stronger
than their pinning by the surface charges, so that the concept of 
an uniform background works exactly.
At $Z \geq 3$ the uniform background is still 
a good approximation for realistic values of the 
radius of closest approach, $b_s$, of counterions and 
discrete surface negative charges. 
On the other hand, this approximation can fail at $Z=1$ because 
all counterions and discrete negative surface charges have a tendency
to form neutral Bjerrum pairs instead of SCL if $b_s \ll R$. 
In this case, $N(0)$ can be small even for $Z=1$.
When negative charges are clustered, for example, 
form compact triplets, 
even at $Z = 3$ interaction with such a cluster 
can be as important as interaction with the neighboring counterions. 
Each counterion temds to neutralize one cluster
forming a neutral dipole. Again, this leads
to stronger binding to the surface and smaller $N(0)$.

This discussion naturally leads us to the third approximation
used above in the calculation of the chemical potential of SCL.
Eq.~(\ref{cpot}) was obtained for point-like counterions.
Actually counterions have a finite size and one wonders 
how this affects these results. 
Our results, of course, make sense only if the counterion radius, $b_c$
is smaller than the radius of a Wigner-Seitz cell, $R$,
so that counterions occupy a small fraction of the surface.
In other words, the idea of SCL or WC works only when objects of large 
charge density arrange themselves on a background
with much smaller charge density.
For treevalent ions on the surface 
with the charge density
$\sigma = 1.0~e/$nm$^{-2}$ the radius of Wigner-Seitz cell $R = 1$~nm, 
so that for a counterion with $b_c = 0.5$~nm
this condition is easily satisfied. 
Positive corrections to the energy per ion
of WC are proportional $(b_c/R)^2$ and appear
due to the fact that the charge 
finite size counterion can not be situated
exactly in the potential minimum created by  
its nearest neighbors. 

Finally, all estimates in this paper 
are based on the use of dielectric constant of water
$D\simeq 80$. For the lateral interactions of counterions near 
the surface of an organic material with a low dielectric constant, the
effective dielectric constant $D$ can be substantially smaller. 
(In a macroscopic approach it is close to $D/2$). 
As a result, absolute values of $|\mu(n,T)|$ 
can grow significantly and $N(0)$ may become even smaller.

\section{Conclusion}

In conclusion, the role of strong lateral 
correlations of Z-valent counterions 
condensed on a charged surface is studied.
It is argued that a strongly correlated liquid (SCL), 
or in other words, a two-dimensional one-component plasma 
is a good model for these correlations.
It is shown that, due
to the additional binding provided by SCL, the concentration of
counterions close to
SCL is exponentially small (see Eq.~(\ref{bc0})). This concentration
depends only on $Z$ and the surface charge density of the macroion
$\sigma$ and serves as a
boundary condition for the Poisson-Boltzmann equation (PBE), which is
still valid far from the surface.
PBE is solved with the boundary condition
(\ref{bc0}) for all three standard geometries.
For a charged cylinder, it is
shown that in the presence of SCL the
Onsager-Manning condensation is
strongly modified. The increasing bulk concentration of 
$Z$-valent counterions, $N(\infty)$, 
makes the net negative charge of the
cylinder smaller than in the Onsager-Manning theory,
drives it through zero and makes it positive. 
Similar changes are predicted for a charged sphere with charge $- Q$.
In this case, charge inversion can result in positive a net charge 
$0.84 \sqrt {QZe}$. All these phenomena happen
while the concentration of $Z : 1$ salt is still 
exponentially small. Technically they follow
from the boundary condition (\ref{bc0}), which in turn is
a result of a strong correlations of counterions of the surface
layer.

This theory can be applied to variety of other
problems. First, one can study more complicated solutions where 
a substantial concentration of 1 : 1 salt is added to Z : 1 salt.
We gave only one example of such a problem in the end of Sec. V. 
Second, this theory should be extended to a finite concentration of 
macroions. In this case problems of a global instability
of such a solution should be addressed, too.
Third, one can use a similar theory for counterions of a larger size
and nonspherical shape~\cite{Shklov98}, provided 
they have larger charge density than the macroion's surface.
For example, a positively charged rigid flat surface can
be screened by a 
solution of a rod like polymer such as double helix DNA~\cite{Yang}.
If projected to the plane, the  negative surface density of 
DNA is by absolute value larger than the charge density of the plane,
DNA rods form strongly correlated nematic liquid, which provides
strong binding of DNA to the surface.
A net charge density of the plane can experience 
correlation induced charge inversion.
These and other problems will be addressed in future publications.

\acknowledgements

I am grateful to V. I. Perel, with whom this work was started,
and to A. Yu. Grosberg, with whom Eq.~(\ref{qs}) was derived 
and who read several drafts of this paper. I acknowledge valuable
discussions with V. A. Bloomfield, M. M. Fogler,
R. Kjellander, S. Marcelja, R. Podgornik, I. Rouzina and M. Voloshin.
This work was supported by NSF DMR-9616880

\end{multicols}
\end{document}